\documentclass[11pt]{article}
\usepackage{amssymb,amsmath,amsfonts}
\usepackage{graphicx}
\usepackage{graphics}
\usepackage{eepic,epsfig}

\begin{document}

\title{Fermionic vacuum densities in higher-dimensional \\
de Sitter spacetime}
\author{E. R. Bezerra de Mello$^{1}$\thanks{%
E-mail: emello@fisica.ufpb.br}\, and A. A. Saharian$^{1,2}$\thanks{%
E-mail: saharian@ictp.it} \\
\\
\textit{$^1$Departamento de F\'{\i}sica-CCEN, Universidade Federal da Para%
\'{\i}ba}\\
\textit{58.059-970, Caixa Postal 5.008, Jo\~{a}o Pessoa, PB, Brazil}\vspace{%
0.3cm}\\
\textit{$^2$Department of Physics, Yerevan State University,}\\
\textit{1 Alex Manoogian Street, 0025 Yerevan, Armenia}}
\maketitle

\begin{abstract}
Fermionic condensate and the vacuum expectation values of the
energy-momentum tensor are investigated for twisted and untwisted massive
spinor fields in higher-dimensional de Sitter spacetime with toroidally
compactified spatial dimensions. The expectation values are presented in the
form of the sum of corresponding quantities in the uncompactified de Sitter
spacetime and the parts induced by non-trivial topology. The latter are
finite and renormalizations are needed for the first parts only. Closed
formulae are derived for the renormalized fermionic vacuum densities in
uncompactified odd-dimensional de Sitter spacetimes. It is shown that,
unlike to the case of 4-dimensional spacetime, for large values of the mass,
these densities are exponentially suppressed. Asymptotic behavior of the
topological parts in the expectation values are investigated in the early
and late stages of the cosmological expansion. When the comoving lengths of
compactified dimensions are much smaller than the de Sitter curvature
radius, the leading term in the topological parts coincide with the
corresponding quantities for a massless fermionic field and are conformally
related to the corresponding flat spacetime results. In this limit the
topological parts dominate the uncompactified de Sitter part and the
back-reaction effects should be taken into account. In the opposite limit,
for a massive field the asymptotic behavior of the topological parts is
damping oscillatory.
\end{abstract}

\bigskip

PACS numbers: 04.62.+v, 04.50.-h, 11.10.Kk, 04.20.Gz

\bigskip

\section{Introduction}

De Sitter (dS) spacetime is one of the simplest and most interesting
spacetimes allowed by general relativity. Quantum field theory in this
background has been extensively studied during the past two decades. Much of
early interest to dS spacetime was motivated by the questions related to the
quantization of fields propagating on curved backgrounds. This spacetime has
a high degree of symmetry and numerous physical problems are exactly
solvable on this background. The importance of this theoretical work
increased by the appearance of \ the inflationary cosmology scenario \cite%
{Lind90}. In most inflationary models, an approximately dS spacetime is
employed to solve a number of problems in standard cosmology. During an
inflationary epoch, quantum fluctuations in the inflaton field introduce
inhomogeneities and may affect the transition toward the true vacuum. These
fluctuations play a central role in the generation of cosmic structures from
inflation. More recently astronomical observations of high redshift
supernovae, galaxy clusters and cosmic microwave background \cite{Ries07}
indicate that at the present epoch, the Universe is accelerating and can be
well approximated by a world with a positive cosmological constant. If the
Universe would accelerate indefinitely, the standard cosmology would lead to
an asymptotic dS universe. Hence, the investigation of physical effects in
dS spacetime is important for understanding both the early Universe and its
future.

Many of high energy theories of fundamental physics are formulated in
higher-dimensional spacetimes. In particular, the idea of extra dimensions
has been extensively used in supergravity and superstring theories. It is
commonly assumed that the extra dimensions are compactified. From an
inflationary point of view, universes with compact spatial dimensions, under
certain conditions, should be considered a rule rather than an exception
\cite{Lind04}. The models of a compact universe with non-trivial topology
may play important roles by providing proper initial conditions for
inflation. As it was argued in Refs. \cite{McIn04}, there is no reason to
believe that the version of dS spacetime which may emerge from string
theory, will necessarily be the most familiar version with symmetry group $%
O(1,4)$ and there are many different topological spaces which can accept the
dS metric locally. There are many reasons to expect that in string theory
the most natural topology for the universe is that of a flat compact
three-manifold. The quantum creation of the universe having toroidal spatial
topology is discussed in \cite{Zeld84} and in references \cite{Gonc85}
within the framework of various supergravity theories.

The compactification of spatial dimensions leads to a number of interesting
quantum field theoretical effects which include instabilities in interacting
field theories \cite{Ford80a}, topological mass generation \cite{Ford79} and
symmetry breaking \cite{Toms80b}. In the case of non-trivial topology, the
boundary conditions imposed on fields give rise to the modification of the
spectrum for vacuum fluctuations and, as a result, to the Casimir-type
contributions in the vacuum expectation values of physical observables (for
the topological Casimir effect and its role in cosmology see \cite%
{Grib94,Most97} and references therein). In the Kaluza-Klein-type models,
the Casimir effect has been used as a stabilization mechanism for moduli
fields which parametrize the size and the shape of the extra dimensions. The
Casimir energy can also serve as a model for dark energy needed for the
explanation of the present accelerated expansion of the universe (see \cite%
{Milt03} and references therein). One-loop quantum effects for various spin
fields on the background of dS spacetime, have been discussed by several
authors (see, for instance, \cite{Cher68}-\cite{Birr82} and references
therein). The effects of the toroidal compactification of spatial dimensions
in dS spacetime on the properties of quantum vacuum for a scalar field with
general curvature coupling parameter are investigated in Refs. \cite%
{Saha07,Bell08} (for quantum effects in braneworld models with dS spaces and
in higher-dimensional brane models with compact internal spaces see, for
instance, Refs. \cite{dSbrane,Flac03}). The one-loop quantum effects for a
fermionic field on background of 4-dimensional dS spacetime with spatial
topology $\mathrm{R}^{p}\times (\mathrm{S}^{1})^{q}$ are studied in \cite%
{Saha08}.

In the present paper, we investigate one-loop quantum effects arising from
vacuum fluctuations of a fermionic field on background of higher-dimensional
dS spacetime with toroidally compactified spatial dimensions. The important
quantities that characterize the quantum fluctuations during the dS
expansion are the fermionic condensate and the expectation value of the
energy-momentum tensor. In the next section, by using the dimensional
regularization procedure, we evaluate these quantities in uncompactified
odd-dimensional dS spacetimes. The plane wave fermionic eigenfunctions in $%
(D+1)$-dimensional dS spacetime with an arbitrary number of toroidally
compactified dimensions are constructed in section \ref{sec:EigFunc}. In
section \ref{sec:FermCond} these eigenfunctions are used for the evaluation
of the fermionic condensate in both cases of the fields with periodicity and
antiperiodicity conditions along compactified dimensions. The behavior of
these quantities are investigated in asymptotic regions of the parameters.
The topological parts in the vacuum expectation values of the
energy-momentum tensor are investigated in section \ref{sec:EMT}. In the
last section we summarize the main results of the paper.

\section{Vacuum expectation values in odd-dimensional \newline
uncompactified dS spacetime}

\label{sec:UncompdS}

In this section we will evaluate the fermionic condensate and the vacuum
expectation value (VEV) of the energy-momentum tensor in odd-dimensional
uncompactified dS spacetimes by using the dimensional regularization
procedure. These quantities are among the most important characteristics of
the vacuum state and carry an information on both local and global
properties. Let us consider a quantum fermionic field $\psi $ on background
of $(D+1)$-dimensional de Sitter spacetime, $\mathrm{dS}_{D+1}$, described
by the line element%
\begin{equation}
ds^{2}=dt^{2}-e^{2t/\alpha }\sum_{i=1}^{D}(dz^{i})^{2},  \label{ds2deSit}
\end{equation}%
where the parameter $\alpha $ in the expression for the scale factor is
related to the corresponding cosmological constant $\Lambda $ by the formula
$\alpha ^{2}=D(D-1)/(2\Lambda )$.

The dynamics of the field in a curved spacetime is governed by the covariant
Dirac equation
\begin{equation}
i\gamma ^{\mu }\nabla _{\mu }\psi -m\psi =0\ ,\;\nabla _{\mu }=\partial
_{\mu }+\Gamma _{\mu },  \label{Direq}
\end{equation}%
where $\gamma ^{\mu }=e_{(a)}^{\mu }\gamma ^{(a)}$ are the generalized Dirac
matrices and $\Gamma _{\mu }$ is the spin connection. The latter is given in
terms of the flat-space Dirac matrices $\gamma ^{(a)}$ by the relation
\begin{equation}
\Gamma _{\mu }=\frac{1}{4}\gamma ^{(a)}\gamma ^{(b)}e_{(a)}^{\nu }e_{(b)\nu
;\mu }\ ,  \label{Gammamu}
\end{equation}%
where the semicolon means the covariant derivative of vector fields. In the
equations above $e_{(a)}^{\mu }$ are the tetrad components defined by $%
e_{(a)}^{\mu }e_{(b)}^{\nu }\eta ^{ab}=g^{\mu \nu }$, with $\eta ^{ab}$
being the Minkowski spacetime metric tensor.

Let $S_{F}(x,x^{\prime })$ be the Feynman Green function for the spinor
field in $(D+1)$-dimensional dS spacetime. In Ref. \cite{Cand75} it has been
shown that for $D<1$ one has the relation%
\begin{equation}
\mathrm{tr}[S_{F}(x,x)]=-\frac{iNm}{(4\pi )^{(D+1)/2}}\frac{\Gamma \left(
\frac{D+1}{2}+im\alpha \right) \Gamma \left( \frac{D+1}{2}-im\alpha \right)
}{\alpha ^{D-1}\Gamma (1-im\alpha )\Gamma (1+im\alpha )}\Gamma \left( \frac{%
1-D}{2}\right) ,  \label{SFtrace}
\end{equation}%
where the trace is taken over the spinor indices of the Green function and $%
N $ is the number of spinor components in $(D+1)$-dimensional spacetime (see
next section). The expression on the right-hand side of this formula is
finite for all even values of $D$. In this case, by using the well-known
properties of the gamma function, formula (\ref{SFtrace}) can also be
written in the form%
\begin{equation}
\mathrm{tr}[S_{F}(x,x)]=-\frac{iNm^{D}}{(4\pi )^{(D+1)/2}}\Gamma \left(
\frac{1-D}{2}\right) \tanh (\pi m\alpha )\prod\limits_{j=0}^{D/2-1}\bigg[1+%
\frac{(j+1/2)^{2}}{m^{2}\alpha ^{2}}\bigg].  \label{SFtrace1}
\end{equation}

To find the renormalized value for the fermionic condensate we should
subtract from (\ref{SFtrace1}) the corresponding DeWitt-Schwinger expansion $%
\mathrm{tr}[S_{F}^{\mathrm{(DS)}}(x,x)]$, truncating the expansion at the
adiabatic order $D$:%
\begin{equation}
\langle \bar{\psi}\psi \rangle _{\mathrm{dS,ren}}=-i\left\{ \mathrm{tr}%
[S_{F}(x,x)]-\mathrm{tr}[^{(D)}S_{F}^{\mathrm{(DS)}}(x,x)]\right\} ,
\label{FCren}
\end{equation}%
where $^{(D)}S_{F}^{\mathrm{(DS)}}(x,x)$ stands for the truncated expansion.
This expansion can be found by using the corresponding result for the
bispinor $G_{F}(x,x^{\prime })$ which is related to the function $%
S_{F}(x,x^{\prime })$ by the formula $S_{F}(x,x^{\prime })=(i\gamma ^{\mu
}\nabla _{\mu }+m)G_{F}(x,x^{\prime })$. The structure of the
DeWitt-Schwinger expansion for the function $G_{F}(x,x^{\prime })$ is
similar to that for the scalar Green function given, for example, in \cite%
{Birr82}:%
\begin{equation}
G_{F}^{\mathrm{(DS)}}(x,x)=\frac{-im^{D-1}}{(4\pi )^{(D+1)/2}}%
\sum_{j=0}^{\infty }\frac{a_{j}(x)}{m^{2j}}\Gamma (j-(D-1)/2),  \label{GFDS1}
\end{equation}%
where the coefficients $a_{j}(x)$ are bispinors. The expressions for the
first three coefficients in terms of the curvature tensor of the background
geometry can be found in \cite{Chri78}. In particular, $a_{0}=I$, $%
a_{1}=-RI/12$, where $I$ is the unit matrix and $R$ is the Ricci scalar. For
de Sitter spacetime the bispinor coefficients $a_{j}(x)$ in the adiabatic
expansion of $G_{F}(x,x^{\prime })$ have the structure $a_{j}(x)=b_{j}R^{j}I$%
, where $b_{j}$ are numerical coefficients and $R=D(D+1)/\alpha ^{2}$ is the
Ricci scalar for $\mathrm{dS}_{D+1}$. For example, the coefficient $a_{2}$
is given by the expression%
\begin{equation}
a_{2}=\frac{D(D+1)}{1440\alpha ^{4}}(D-2)(5D+7)I.  \label{a2}
\end{equation}

Now, by using the relation $\mathrm{tr}[\gamma ^{\mu }]=0$ for the Dirac
matrices, we see that for dS spacetime one has the relation $\mathrm{tr}%
[S_{F}^{\mathrm{(DS)}}(x,x)]=m\mathrm{tr}[G_{F}^{\mathrm{(DS)}}(x,x)]$.
Hence, by making use of formulae (\ref{FCren}) and (\ref{GFDS1}), for the
renormalized value of the fermionic condensate we find:%
\begin{eqnarray}
\langle \bar{\psi}\psi \rangle _{\mathrm{dS,ren}} &=&-\frac{Nm^{D}}{(4\pi
)^{(D+1)/2}}\Gamma \left( \frac{1-D}{2}\right) \bigg\{\tanh (\pi m\alpha
)\prod\limits_{j=0}^{D/2-1}\bigg[1+\frac{(j+1/2)^{2}}{m^{2}\alpha ^{2}}\bigg]
\notag \\
&&-1-\sum_{j=1}^{D/2}a_{j}(x)\frac{(1-D)\cdots (2l-1-D)}{2^{j}m^{2j}}\bigg\}.
\label{FCren1}
\end{eqnarray}%
Further, we present the hyperbolic tangent function in this formula in the
form $\tanh (x)=1-2(e^{2x}+1)^{-1}$. The product in curly braces of (\ref%
{FCren1}) multiplied by the first term in this representation is cancelled
by the adiabatic subtraction terms\footnote{%
We have explicitly checked this for the terms involving $(m\alpha )^{-2}$
and $(m\alpha )^{-4}$; in the case $D=5$ these include all adiabatic terms.}
and the renormalized fermionic condensate in odd-dimensional dS spacetimes
is presented in the form%
\begin{equation}
\langle \bar{\psi}\psi \rangle _{\mathrm{dS,ren}}=\frac{2Nm^{D}}{(4\pi
)^{(D+1)/2}}\frac{\Gamma \left( (1-D)/2\right) }{e^{2\pi m\alpha }+1}%
\prod\limits_{j=0}^{D/2-1}\bigg[1+\frac{(j+1/2)^{2}}{m^{2}\alpha ^{2}}\bigg].
\label{FermConddSRen}
\end{equation}%
Note that the sign of the fermionic condensate is determined by the sign of
the gamma function. For a massless fermionic field from (\ref{FermConddSRen}%
) we find the following result:%
\begin{equation}
\langle \bar{\psi}\psi \rangle _{\mathrm{dS,ren}}=\frac{(-1)^{D/2}N}{(4\pi
)^{(D+1)/2}\alpha ^{D}}\Gamma \left( \frac{D+1}{2}\right) .  \label{FCm0}
\end{equation}

\begin{figure}[tbph]
\begin{center}
\epsfig{figure=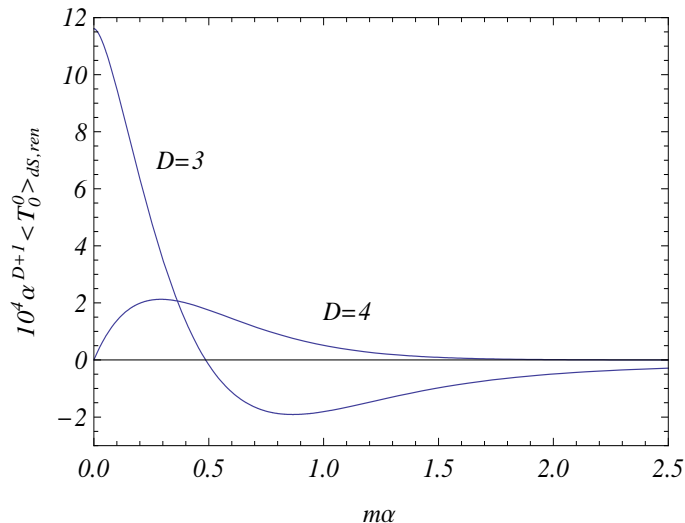,width=9.cm,height=7.cm}
\end{center}
\caption{The vacuum energy density in uncompactified dS spacetime, $\langle
T_{0}^{0}\rangle _{\mathrm{dS,ren}}$, as a function of the parameter $m%
\protect\alpha $ for spatial dimensions $D=3,4$. }
\label{fig1}
\end{figure}

Having the fermionic condensate, we can evaluate the VEV of the
energy-momentum tensor for the fermionic field in odd-dimensional dS
spacetime by using the trace relation%
\begin{equation}
\langle T_{l}^{l}\rangle _{\mathrm{dS,ren}}=m\langle \bar{\psi}\psi \rangle
_{\mathrm{dS,ren}}.  \label{TraceRel1}
\end{equation}%
Since the dS spacetime is maximally symmetric, we have $\langle
T_{l}^{k}\rangle _{\mathrm{dS,ren}}=\mathrm{const}\,\delta _{l}^{k}$. The
constant in this relation is found from (\ref{TraceRel1}). In this way, for
even values $D$, for the renormalized VEV of the energy-momentum tensor we
find:%
\begin{equation}
\langle T_{l}^{k}\rangle _{\mathrm{dS,ren}}=\frac{2Nm^{D+1}\delta _{l}^{k}}{%
(4\pi )^{(D+1)/2}(D+1)}\frac{\Gamma \left( (1-D)/2\right) }{e^{2\pi m\alpha
}+1}\prod\limits_{j=0}^{D/2-1}\bigg[1+\frac{(j+1/2)^{2}}{m^{2}\alpha ^{2}}%
\bigg].  \label{TkldSren}
\end{equation}%
For a massless fermionic field this tensor vanishes. Of course, we could
expect this result, since in odd-dimensional spacetimes the trace anomaly is
absent. For large values of the mass, $m\alpha \gg 1$, the VEVs for both
fermionic condensate and the energy-momentum tensor are exponentially
suppressed. In particular, for 5-dimensional de Sitter spacetime formula (%
\ref{TkldSren}) leads to the result%
\begin{equation}
\langle T_{l}^{k}\rangle _{\mathrm{dS,ren}}=\frac{m^{5}\delta _{l}^{k}}{%
15\pi ^{2}}\frac{1}{e^{2\pi m\alpha }+1}\left( 1+\frac{5}{2m^{2}\alpha ^{2}}+%
\frac{9}{16m^{4}\alpha ^{4}}\right) ,\;D=4.  \label{TDS5}
\end{equation}%
For $D=3$ the renormalized VEV of the energy-momentum tensor for fermionic
field in dS spacetime is investigated in \cite{Mama81} (see also \cite%
{Grib94}). The corresponding expression has the form%
\begin{equation}
\langle T_{l}^{k}\rangle _{\mathrm{dS,ren}}^{(D=3)}=\frac{\delta _{l}^{k}}{%
16\pi ^{2}\alpha ^{4}}\left\{ 2m^{2}\alpha ^{2}(m^{2}\alpha ^{2}+1)[\ln
(m\alpha )-{\mathrm{Re\,}}\psi (im\alpha )]+m^{2}\alpha ^{2}/6+11/60\right\}
,  \label{TkldSren3}
\end{equation}%
where $\psi (x)$ is the logarithmic derivative of the gamma-function. In the
limit of zero mass, only the last term in curly braces contributes. For
large values of the mass we have:%
\begin{equation}
\langle T_{l}^{k}\rangle _{\mathrm{dS,ren}}^{(D=3)}\approx -\frac{\delta
_{l}^{k}}{960\pi ^{2}\alpha ^{6}m^{2}},\;m\alpha \gg 1,  \label{largeMassD3}
\end{equation}%
and, unlike to the case of odd dimensions, the suppression is power-law. In
figure \ref{fig1} we have plotted the vacuum energy density $\langle
T_{0}^{0}\rangle _{\mathrm{dS,ren}}$ as a function of the parameter $m\alpha
$ for spatial dimensions $D=3,4$.

In dS spacetime the energy-momentum tensor $\langle T_{l}^{k}\rangle _{%
\mathrm{dS,ren}}$ corresponds to a gravitational source of the cosmological
constant type. Combining with the initial cosmological constant $\Lambda $,
one-loop effects in uncompactified dS spacetime lead to the effective
cosmological constant
\begin{equation}
\Lambda _{\mathrm{eff}}=\Lambda +8\pi G_{D+1}\langle T_{0}^{0}\rangle _{%
\mathrm{dS,ren}},  \label{effCC}
\end{equation}%
where $G_{D+1}$ is the gravitational constant in $(D+1)$-dimensional
spacetime. In the discussion above we have assumed that the quantum state of
a fermionic field is the Bunch-Davies vacuum state (also called Euclidean
vacuum). In \cite{Ande00} it was shown that for a scalar field with a wide
range of mass and curvature coupling parameter the expectation values of the
energy-momentum tensor in arbitrary physically admissable states approaches
the expectation value in the Bunch-Davies vacuum at late times.

\section{Plane wave eigenspinors in toroidally compactified de Sitter
spacetime}

\label{sec:EigFunc}

Now we turn to the investigation of one-loop quantum effects induced by the
compactness of the spatial dimensions in dS spacetime. We consider $(D+1)$%
-dimensional dS spacetime with spatial topology $\mathrm{R}^{p}\times (%
\mathrm{S}^{1})^{q}$, $p+q=D$. The compactification of the spatial
dimensions leads to the modification of the spectrum for zero-point
fluctuations of fields and as a result of this the VEVs of physical
observables are changed. This is the topological Casimir effect which is
well-investigated in the literature for various type of geometries.

In the $(D+1)$-dimensional flat spacetime the Dirac matrices are $N\times N$
matrices with $N=2^{[(D+1)/2]}$ (see, for instance, \cite{Pais62}), where
the square brackets mean the integer part of the enclosed expression. In the
discussion below we will assume that these matrices are given in the chiral
representation:
\begin{equation}
\gamma ^{(0)}=\left(
\begin{array}{cc}
1 & 0 \\
0 & -1%
\end{array}%
\right) ,\;\gamma ^{(a)}=\left(
\begin{array}{cc}
0 & \sigma _{a} \\
-\sigma _{a}^{+} & 0%
\end{array}%
\right) ,\;a=1,2,\ldots ,D.  \label{gam0l}
\end{equation}%
From the anticommutation relations for the Dirac matrices one has $\sigma
_{a}\sigma _{b}^{+}+\sigma _{b}\sigma _{a}^{+}=2\delta _{ab}$. For example,
in $D=4$ the first four matrices $\gamma ^{(a)}$, $a=0,1,2,3$, can be taken
the same as the corresponding matrices in 4-dimensional spacetime and $%
\gamma ^{(4)}=\gamma ^{(0)}\gamma ^{(1)}\gamma ^{(2)}\gamma ^{(3)}$. In this
case $\sigma _{1},\sigma _{2},\sigma _{3}$ are the standard Pauli matrices
and%
\begin{equation}
\sigma _{4}=\left(
\begin{array}{cc}
0 & -i \\
-i & 0%
\end{array}%
\right) .  \label{sigma4}
\end{equation}%
Note that, unlike to the Pauli matrices, $\sigma _{4}$ is antihermitian.

In order to evaluate the VEVs of the fermionic condensate and the
energy-momentum tensor in toroidally compactified dS spacetime, we will use
the direct mode-summation procedure. In this procedure we need to know a
complete set of properly normalized eigenspinors $\{\psi _{\beta
}^{(+)},\psi _{\beta }^{(-)}\}$, specified by the collective index $\beta $.
By virtue of spatial translation invariance the spatial part of the
eigenfunctions $\psi _{\beta }^{(\pm )}$ can be taken in the standard plane
wave form $e^{\pm i\mathbf{kz}}$, where $\mathbf{k}$ is the wave vector. We
will decompose the vectors $\mathbf{z}$ and $\mathbf{k}$ into the components
along the uncompactified and compactified dimensions, $\mathbf{z=}(\mathbf{z}%
_{p},\mathbf{z}_{q})$, $\mathbf{k=}(\mathbf{k}_{p},\mathbf{k}_{q})$. One of
the characteristic features of the field theory on backgrounds with
non-trivial topology is the appearance of inequivalent types of fields with
the same spin \cite{Isha78}. For fermion fields the boundary conditions
along the compactified dimensions can be either periodic (untwisted field)
or antiperiodic (twisted field). First we consider the field with
periodicity conditions (no summation over $l$):%
\begin{equation}
\psi (t,\mathbf{z}_{p},\mathbf{z}_{q}+L_{l}\mathbf{e}_{l})=\psi (t,\mathbf{z}%
_{p},\mathbf{z}_{q}),  \label{bc}
\end{equation}%
where $\mathbf{e}_{l}$\ is the unit vector in the direction of the
coordinate $z^{l}$ with the length $L_{l}$, $0\leqslant z^{l}\leqslant L_{l}$%
. The case of a fermionic field with antiperiodicity conditions will be
discussed below. For a spinor field with periodicity conditions along the
compactified dimensions the corresponding wave vector has the components%
\begin{equation}
\mathbf{k}_{q}=(2\pi n_{p+1}/L_{p+1},\ldots ,2\pi n_{D}/L_{D}),  \label{kq}
\end{equation}%
where $n_{p+1},\ldots ,n_{D}=0,\pm 1,\pm 2,\ldots $.

In order to find the time dependence of the fermionic eigenfunctions, we
choose the basis tetrad in the form $e_{\mu }^{(0)}=\delta _{\mu }^{0}$,$%
\;e_{\mu }^{(a)}=e^{t/\alpha }\delta _{\mu }^{a}$,$\;a=1,2,\ldots ,D$. For
the components of the corresponding spin connection we have $\Gamma _{0}=0$,$%
\;\Gamma _{l}=(e^{t/\alpha }/2\alpha )\gamma ^{(0)}\gamma ^{(l)}$,$%
\;l=1,2,\ldots ,D$. Decomposing the $N$-component spinor fields in the form%
\begin{equation}
\psi _{\beta }^{(+)}=e^{i\mathbf{kz}}\left(
\begin{array}{c}
\varphi _{\beta ,+}^{(+)} \\
\varphi _{\beta ,-}^{(+)}%
\end{array}%
\right) ,  \label{psidecomp}
\end{equation}%
from the Dirac equation one finds the set of two first order differential
equations:%
\begin{eqnarray}
\left( \eta \partial _{\eta }-D/2-i\alpha m\right) \varphi _{\beta
,+}^{(+)}-i\eta (\mathbf{k}\cdot \boldsymbol{\sigma })\varphi _{\beta
,-}^{(+)} &=&0,  \label{xieq2} \\
\left( \eta \partial _{\eta }-D/2+i\alpha m\right) \varphi _{\beta
,-}^{(+)}-i\eta (\mathbf{k}\cdot \boldsymbol{\sigma }^{+})\varphi _{\beta
,+}^{(+)} &=&0,  \label{xieq3}
\end{eqnarray}%
where $\boldsymbol{\sigma }=(\sigma _{1},\sigma _{2},\ldots ,\sigma _{D})$.
In these equations we have introduced a new independent variable $\eta $ in
accordance with%
\begin{equation}
\eta =\alpha e^{-t/\alpha },\;0\leqslant \eta <\infty .  \label{etavar}
\end{equation}%
Note that $\tau =-\eta $ is the conformal time coordinate, in terms of which
the dS line element (\ref{ds2deSit}) takes the conformally flat form.

By using the relation $(\mathbf{k}\cdot \boldsymbol{\sigma })(\mathbf{k}%
\cdot \boldsymbol{\sigma }^{+})=k^{2}$, with
\begin{equation}
k=\sqrt{\mathbf{k}_{p}^{2}+\mathbf{k}_{q}^{2}},  \label{ka}
\end{equation}%
from (\ref{xieq2}) we obtain the second order differential equations for the
upper and lower components of the spinor:%
\begin{equation}
\lbrack \eta ^{2}\partial _{\eta }^{2}-D\eta \partial _{\eta }+k^{2}\eta
^{2}+\alpha ^{2}m^{2}\pm i\alpha m+D(D+2)/4]\varphi _{\beta ,\pm }^{(+)}=0.
\label{xieq4}
\end{equation}%
These equations are solved in terms of the Hankel functions and the
corresponding solutions have the form%
\begin{equation}
\varphi _{\beta ,\pm }^{(+)}=\eta ^{(D+1)/2}\left[ C_{1,\pm }H_{\pm
1/2-i\alpha m}^{(1)}(k\eta )+C_{2,\pm }H_{\pm 1/2-i\alpha m}^{(2)}(k\eta )%
\right] .  \label{phibetComb}
\end{equation}%
Different choices of the coefficients $C_{j,\pm }$ correspond to different
vacuum states. As in section \ref{sec:UncompdS}, we will assume that the
fermionic field is prepared in the de Sitter invariant Bunch-Davies vacuum,
for which $C_{2,\pm }=0$. Hence, for a fermionic field in the Bunch-Davies
vacuum state the solutions to (\ref{xieq4}) are the functions%
\begin{eqnarray}
\varphi _{\beta ,+}^{(+)} &=&\varphi ^{(c)}\eta ^{(D+1)/2}H_{1/2-i\alpha
m}^{(1)}(k\eta ),\;  \notag \\
\varphi _{\beta ,-}^{(+)} &=&-i\varphi ^{(c)}(\mathbf{n}\cdot %
\boldsymbol{\sigma })\eta ^{(D+1)/2}H_{-1/2-i\alpha m}^{(1)}(k\eta ),
\label{phixiH2}
\end{eqnarray}%
where $\varphi ^{(c)}$ is an arbitrary constant spinor and $\mathbf{n}=%
\mathbf{k}/k$. Note that, in order to obtain the relation between the
coefficients in (\ref{phixiH2}) we have used equation (\ref{xieq2}).

Now, we can construct the positive frequency solutions to the Dirac equation
on the base of functions (\ref{phixiH2}):
\begin{equation}
\psi _{\beta }^{(+)}=A_{\beta }\eta ^{(D+1)/2}e^{i\mathbf{k}\cdot \mathbf{r}%
}\left(
\begin{array}{c}
H_{1/2-i\alpha m}^{(1)}(k\eta )w_{\sigma }^{(+)} \\
-i(\mathbf{n}\cdot \boldsymbol{\sigma })H_{-1/2-i\alpha m}^{(1)}(k\eta
)w_{\sigma }^{(+)}%
\end{array}%
\right) ,  \label{psibet+n}
\end{equation}%
where $\beta =(\mathbf{k},\sigma )$, and $w_{\sigma }^{(+)}$, $\sigma
=1,\ldots ,N/2$, are one-column matrices having $N/2$ rows with the elements
$w_{l}^{(\sigma )}=\delta _{l\sigma }$. In the similar way, for the negative
frequency solutions we find
\begin{equation}
\psi _{\beta }^{(-)}=A_{\beta }\eta ^{(D+1)/2}e^{-i\mathbf{k}\cdot \mathbf{r}%
}\left(
\begin{array}{c}
i(\mathbf{n}\cdot \boldsymbol{\sigma })H_{-1/2+i\alpha m}^{(2)}(k\eta
)w_{\sigma }^{(-)} \\
H_{1/2+i\alpha m}^{(2)}(k\eta )w_{\sigma }^{(-)}%
\end{array}%
\right) ,  \label{psibet-}
\end{equation}%
with $w_{\sigma }^{(-)}=iw_{\sigma }^{(+)}$.

From the orthonormalization condition for the eigenfunctions (\ref{psibet+n}%
) and (\ref{psibet-}), for the coefficient $A_{\beta }$\ one obtains%
\begin{equation}
A_{\beta }^{2}=\frac{ke^{\pi \alpha m}}{2^{p+2}\pi ^{p-1}V_{q}\alpha ^{D}},
\label{Abet}
\end{equation}%
where $V_{q}=L_{p+1}\cdots L_{D}$ is the volume of the compactified
subspace. For a massless fermionic field the Hankel functions are expressed
in terms of exponentials ones and we have the standard conformal relation $%
\psi _{\beta }^{(\pm )}=(\eta /\alpha )^{(D+1)/2}\psi _{\beta }^{\mathrm{(M)}%
(\pm )}$ between eigenspinors (\ref{psibet+n}) and (\ref{psibet-}) defining
the Bunch-Davies vacuum in dS spacetime and the corresponding eigenspinors $%
\psi _{\beta }^{\mathrm{(M)}(\pm )}$ for the Minkowski spacetime with
spatial topology $\mathrm{R}^{p}\times (\mathrm{S}^{1})^{q}$.

The plane wave eigenspinors for a twisted spinor field are constructed in a
similar way. For this field we have the antiperiodicity conditions along the
compactified dimensions:%
\begin{equation}
\psi (t,\mathbf{z}_{p},\mathbf{z}_{q}+L_{l}\mathbf{e}_{l})=-\psi (t,\mathbf{z%
}_{p},\mathbf{z}_{q}).  \label{AntCond}
\end{equation}%
The corresponding eigenfunctions are given by formulae (\ref{psibet+n}), (%
\ref{psibet-}), where now the components of the wave vector along the
compactified dimensions are given by the formula
\begin{equation}
\mathbf{k}_{q}=(\pi (2n_{p+1}+1)/L_{p+1},\ldots ,\pi (2n_{D}+1)/L_{D}),
\label{kqCompTw}
\end{equation}%
with $n_{p+1},\ldots ,n_{D}=0,\pm 1,\pm 2,\ldots $, and
\begin{equation}
k^{2}=k_{p}^{2}+\sum_{l=p+1}^{D}\left[ \pi (2n_{l}+1)/L_{l}\right] ^{2}.
\label{kTw}
\end{equation}%
Note that the physical wave vector is given by the combination $\mathbf{k}%
\eta /\alpha $.

\section{Fermionic condensate in toroidally compactified \newline
de Sitter spacetime}

\label{sec:FermCond}

In this section and in the next one, we will investigate the fermionic
vacuum effects induced by the non-trivial topology of the dS spacetime. The
scheme of the corresponding calculations is similar to that for the case of $%
\mathrm{dS}_{4}$ given in \cite{Saha08}, and we will omit the details.
Expanding the field operator in terms of the complete set of eigenfunctions,
described in the previous section, and using the commutation relations
between the fields operator, we find the mode-sum formula for the fermionic
condensate:
\begin{equation}
\langle \bar{\psi}\psi \rangle _{p,q}=\int d\mathbf{k}_{p}\sum_{\mathbf{k}%
_{q},\sigma }\bar{\psi}_{\mathbf{k},\sigma }^{(-)}\psi _{\mathbf{k},\sigma
}^{(-)},  \label{modesumCond}
\end{equation}%
where $\bar{\psi}_{\mathbf{k},\sigma }^{(-)}=\psi _{\mathbf{k},\sigma
}^{(-)+}\gamma ^{0}$ is the Dirac adjoint. Applying to the sum over $n_{p+1}$
in (\ref{modesumCond}) the Abel-Plana formula (see, for instance, \cite%
{Saha07Gen}), we find the following recurrence formula for the fermionic
condensate in dS spacetime with spatial topology $\mathrm{R}^{p}\times (%
\mathrm{S}^{1})^{q}$:%
\begin{equation}
\langle \bar{\psi}\psi \rangle _{p,q}=\langle \bar{\psi}\psi \rangle
_{p+1,q-1}+\Delta _{p+1}\langle \bar{\psi}\psi \rangle _{p,q},
\label{fermconddec}
\end{equation}%
where $\langle \bar{\psi}\psi \rangle _{p+1,q-1}$ is the fermionic
condensate in dS spacetime having topology $\mathrm{R}^{p+1}\times (\mathrm{S%
}^{1})^{q-1}$, and the term%
\begin{eqnarray}
\Delta _{p+1}\langle \bar{\psi}\psi \rangle _{p,q} &=&\frac{8N\eta
^{D+1}\alpha ^{-D}L_{p+1}}{(2\pi )^{(p+3)/2}V_{q}}\sum_{n=1}^{\infty
}\sum_{n_{p+2}=-\infty }^{+\infty }\cdots \sum_{n_{D}=-\infty }^{+\infty
}\int_{0}^{\infty }dx\,x^{2}  \notag \\
&&\times \frac{{\mathrm{Im}}\left[ K_{1/2-i\alpha m}(\eta x)I_{1/2+i\alpha
m}(\eta x)\right] }{(nL_{p+1})^{p-1}}f_{(p-1)/2}(nL_{p+1}\sqrt{x^{2}+k_{%
\mathbf{n}_{q-1}}^{2}}),  \label{DeltCond2}
\end{eqnarray}%
is induced by the compactness of the $(p+1)$th dimension. In Eq. (\ref%
{DeltCond2}), $I_{\nu }(x)$ and $K_{\nu }(x)$ are the modified Bessel
functions and we have introduced the notations%
\begin{equation}
f_{\nu }(x)\equiv x^{\nu }K_{\nu }(x),\;k_{\mathbf{n}_{q-1}}^{2}=%
\sum_{l=p+2}^{D}(2\pi n_{l}/L_{l})^{2}.  \label{kndmin1}
\end{equation}%
Note that for a massless field the topological part in the fermionic
condensate vanishes.

On the base of the recurrence relation (\ref{fermconddec}) the fermionic
condensate can be decomposed as%
\begin{equation}
\langle \bar{\psi}\psi \rangle _{p,q}=\langle \bar{\psi}\psi \rangle _{%
\mathrm{dS,ren}}+\langle \bar{\psi}\psi \rangle _{\mathrm{c}},\;\langle \bar{%
\psi}\psi \rangle _{\mathrm{c}}=\sum_{l=1}^{q}\Delta _{D+1-l}\langle \bar{%
\psi}\psi \rangle _{D-l,l},  \label{ConddS}
\end{equation}%
where the fermionic condensate in uncompactified dS spacetime, $\langle \bar{%
\psi}\psi \rangle _{\mathrm{dS,ren}}$, is given by formula (\ref%
{FermConddSRen}) and the part $\langle \bar{\psi}\psi \rangle _{\mathrm{c}}$
is induced by the non-trivial spatial topology. Note that the expression on
the right of formula (\ref{DeltCond2}) is finite and the renormalization is
needed for the uncompactified dS part only. Of course, we could expect the
finiteness of the topological part, since the toroidal compactification does
not change the local geometry and, hence, the structure of the divergences
is the same as in uncompactified dS spacetime. From formula (\ref{DeltCond2}%
) we see that the topological part depends on the variable $\eta $ and the
length scales $L_{l}$ in the combinations $L_{l}/\eta $. Noting that $a(\eta
)L_{l}$ is the comoving length with $a(\eta )=\alpha /\eta $ being the scale
factor, we conclude that the topological part of the fermionic condensate is
a function of the comoving lengths of the compactified dimensions.

The general formula for the topological part in the fermionic condensate is
simplified in the asymptotic regions of the parameters. In the limit when
the comoving length of the $(p+1)$th dimension is much smaller than the dS
curvature radius, $a(\eta )L_{p+1}\ll \alpha $, the leading term in the
asymptotic expansion of the fermionic condensate has the form%
\begin{equation}
\Delta _{p+1}\langle \bar{\psi}\psi \rangle _{p,q}\approx -\frac{2N(\eta
/\alpha )^{D-1}m}{(2\pi )^{p/2+1}L_{p+1}^{p-1}V_{q}}\sum_{n=1}^{\infty
}\sum_{n_{p+2}=-\infty }^{+\infty }\cdots \sum_{n_{D}=-\infty }^{+\infty }%
\frac{f_{p/2}(nL_{p+1}k_{\mathbf{n}_{q-1}})}{n^{p}}.  \label{CondEarly}
\end{equation}%
In this limit the topological part is negative. Taking into account the
relation between the conformal and synchronous time coordinates, we see that
formula (\ref{CondEarly}) describes the asymptotic behavior in the early
stages of the cosmological expansion corresponding to $t\rightarrow -\infty $%
. Since the part $\langle \bar{\psi}\psi \rangle _{\mathrm{dS,ren}}$ in the
fermionic condensate is time-independent, we conclude that in this limit the
topological part dominates.

Now let us consider large values of the comoving compactification scale, $%
a(\eta )L_{p+1}\gg \alpha $. In terms of the synchronous time coordinate
this corresponds to the late stages of the cosmological evolution, $%
t\rightarrow +\infty $. By using the formulae for the modified Bessel
functions for small values of the arguments, we find the leading term given
below:%
\begin{equation}
\Delta _{p+1}\langle \bar{\psi}\psi \rangle _{p,q}\approx -\frac{N\alpha
B_{0}\sin [2mt-2\alpha m\ln (\alpha /L_{p+1})-\phi _{0}]}{2^{p/2-1}\pi
^{(p+1)/2}L_{p+1}^{p+1}V_{q}\cosh (\alpha m\pi )e^{(D+1)t/\alpha }}.
\label{CondLate}
\end{equation}%
The constants in the expression on the right-hand side are defined by the
relation%
\begin{equation}
B_{0}e^{i\phi _{0}}=\frac{2^{-i\alpha m}}{\Gamma (1/2+i\alpha m)}%
\sum_{n=1}^{\infty }\sum_{n_{p+2}=-\infty }^{+\infty }\cdots
\sum_{n_{D}=-\infty }^{+\infty }\frac{f_{p/2+1+i\alpha m}(nL_{p+1}k_{\mathbf{%
n}_{q-1}})}{n^{p+2+2i\alpha m}}.  \label{B0}
\end{equation}%
Hence, in the limit when the comoving length of the compactified dimensions
is much larger than the curvature radius of dS spacetime, for a massive
fermionic field the topological part oscillates with the amplitude
exponentially decreasing with respect to the synchronous time coordinate.
The damping factor in the amplitude and the oscillation frequency are the
same for all terms in the sum over $l$ in the expression (\ref{ConddS}) for $%
\langle \bar{\psi}\psi \rangle _{\mathrm{c}}$ and we have the similar
oscillating behavior for the total topological term: $\langle \bar{\psi}\psi
\rangle _{c}\propto e^{-(D+1)t/\alpha }\sin \left( 2mt+\phi _{\mathrm{c}%
}\right) $.

In the case of a fermionic field with antiperiodicity conditions (\ref%
{AntCond}), the fermionic condensate can be found in a way similar to that
for the periodicity conditions. It can be seen that the corresponding
formulae for the topological parts are obtained from those for the field
with periodicity conditions inserting the factor $(-1)^{n}$ in the summation
over $n$ and replacing the definition for $k_{\mathbf{n}_{q-1}}^{2}$ by
\begin{equation}
k_{\mathbf{n}_{q-1}}^{2}=\sum_{l=p+2}^{D}\left[ \pi (2n_{l}+1)/L_{l}\right]
^{2}.  \label{knq-1Tw}
\end{equation}%
In situations where the main contribution comes from the term $n=1$, the
topological parts in the fermionic condensate for fields with periodicity
and antiperiodicity conditions have opposite signs.

\section{Vacuum expectation value of the energy-momentum tensor}

\label{sec:EMT}

The one-loop topological effects in the VEV of the energy-momentum tensor
are investigated in a way similar to that for the fermionic condensate. The
corresponding mode-sum has the form%
\begin{equation}
\langle T_{\mu \nu }\rangle =\frac{i}{2}\int d\mathbf{k}_{p}\sum_{\mathbf{k}%
_{q},\sigma }[\bar{\psi}_{\mathbf{k},\sigma }^{(-)}\gamma _{(\mu }\nabla
_{\nu )}\psi _{\mathbf{k},\sigma }^{(-)}-(\nabla _{(\mu }\bar{\psi}_{\mathbf{%
k},\sigma }^{(-)})\gamma _{\nu )}\psi _{\mathbf{k},\sigma }^{(-)}].
\label{VEVEMT}
\end{equation}%
For a fermionic field obeying periodic boundary conditions, we substitute
the eigenfunctions given in (\ref{psibet-}) into this formula and by using
the Abel-Plana formula, the VEV of the energy-momentum tensor in dS
spacetime with spatial topology $\mathrm{R}^{p}\times (\mathrm{S}^{1})^{q}$
is presented in the decomposed form%
\begin{equation}
\langle T_{k}^{l}\rangle _{p,q}=\langle T_{k}^{l}\rangle _{p+1,q-1}+\Delta
_{p+1}\langle T_{k}^{l}\rangle _{p,q}.  \label{TllDecomp}
\end{equation}%
Here $\langle T_{k}^{l}\rangle _{p+1,q-1}$ is the VEV of the energy-momentum
tensor for the topology $\mathrm{R}^{p+1}\times (\mathrm{S}^{1})^{q-1}$ and
the part (no summation over $l$)%
\begin{eqnarray}
\Delta _{p+1}\langle T_{k}^{l}\rangle _{p,q} &=&\frac{N\eta ^{D+2}\alpha
^{-D-1}L_{p+1}\delta _{k}^{l}}{(2\pi )^{(p+1)/2}V_{q}}\sum_{n=1}^{\infty
}\sum_{n_{p+2}=-\infty }^{+\infty }\cdots \sum_{n_{D}=-\infty }^{+\infty
}\int_{0}^{\infty }dx\,x  \notag \\
&&\times \frac{{\mathrm{Re}}[I_{-1/2-i\alpha m}^{2}(\eta x)-I_{1/2+i\alpha
m}^{2}(\eta x)]}{\cosh (\alpha m\pi )(L_{p+1}n)^{p+1}}f_{p}^{(l)}(nL_{p+1}%
\sqrt{x^{2}+k_{\mathbf{n}_{q-1}}^{2}}),  \label{TopTll}
\end{eqnarray}%
is due to the compactness of the $(p+1)$th dimension. For the separate
components the functions $f_{p}^{(l)}(y)$ have the form
\begin{eqnarray}
f_{p}^{(l)}(y) &=&f_{(p+1)/2}(y),\;l=0,1,\ldots ,p,  \notag \\
f_{p}^{(p+1)}(y) &=&-pf_{(p+1)/2}(y)-y^{2}f_{(p-1)/2}(y),\;\;  \label{fpl} \\
f_{p}^{(l)}(y) &=&k_{l}^{2}(nL_{p+1})^{2}f_{(p-1)/2}(y),\;l=p+2,\ldots ,D,
\notag
\end{eqnarray}%
where $k_{l}=2\pi n_{l}/L_{l}$ and the function $f_{\nu }(x)$ is defined in (%
\ref{kndmin1}). The topological parts (\ref{TopTll}) are finite
and by the recurrence relation (\ref{TllDecomp}) the
renormalization procedure is reduced to the renormalization of the
corresponding VEV in uncompactified dS spacetime. As we have
mentioned before, this property is a consequence of the fact that
the toroidal compactification of the dS spacetime considered in
the present paper does not change the local geometry and the
structure of the divergences is the same as in the uncompactified
dS spacetime. Another important point to be mentioned is related
to the finite renormalization terms in the VEVs. These terms are
completely determined by the local geometry and in our
compactification scheme they do not depend on the parameters
characterizing the non-trivial topology (lengths of the
compactified dimensions). As a result the topological parts are
not touched by the finite renormalization terms. It can be seen
that the topological part satisfies the trace relation $\Delta
_{p+1}\langle T_{l}^{l}\rangle _{p,q}=m\Delta _{p+1}\langle
\bar{\psi}\psi \rangle _{p,q}$ and is covariantly conserved in dS
spacetime: $\left( \Delta _{p+1}\langle T_{l}^{k}\rangle
_{p,q}\right) _{;k}=0$. In particular, this part is traceless for
a massless field. In the uncompactified subspace the equation of
state for the topological part of the energy-momentum tensor is of
the cosmological constant type. Note that the topological parts
are time-dependent and they break the dS symmetry.

After the repetitive application of the recurrence formula (\ref{TllDecomp}%
), the VEV of the energy-momentum tensor for the topology $\mathrm{R}%
^{p}\times (\mathrm{S}^{1})^{q}$ is presented in the form%
\begin{equation}
\langle T_{l}^{k}\rangle _{p,q}=\langle T_{l}^{k}\rangle _{\mathrm{dS,ren}%
}+\langle T_{l}^{k}\rangle _{\mathrm{c}},\;\langle T_{l}^{k}\rangle _{%
\mathrm{c}}=\sum_{l=1}^{q}\Delta _{D+1-l}\langle T_{l}^{k}\rangle _{D-l,l},
\label{TlkdS}
\end{equation}%
where the renormalized VEV in the uncompactified dS spacetime, $\langle
T_{l}^{k}\rangle _{\mathrm{dS,ren}}$, is given by formula (\ref{TkldSren})
and $\langle T_{l}^{k}\rangle _{\mathrm{c}}$ is the topological part.

For a massless fermionic field, from (\ref{TopTll}) one finds (no summation
over $l$)%
\begin{equation}
\Delta _{p+1}\langle T_{l}^{l}\rangle _{p,q}=\frac{2N(\eta /\alpha )^{D+1}}{%
(2\pi )^{p/2+1}V_{q}L_{p+1}^{p+1}}\sum_{n=1}^{\infty }\sum_{n_{p+2}=-\infty
}^{+\infty }\cdots \sum_{n_{D}=-\infty }^{+\infty }\frac{%
g_{p}^{(l)}(nL_{p+1}k_{\mathbf{n}_{q-1}})}{n^{p+2}},  \label{DelTConf}
\end{equation}%
with the notations%
\begin{eqnarray}
g_{p}^{(l)}(y) &=&f_{p/2+1}(y),\;l=0,1,\ldots ,p,  \notag \\
g_{p}^{(p+1)}(y) &=&-(p+1)f_{p/2+1}(y)-y^{2}f_{p/2}(y),  \label{gi} \\
g_{p}^{(l)}(y) &=&(nL_{p+1}k_{l})^{2}f_{p/2}(y),\;l=p+2,\ldots ,D.  \notag
\end{eqnarray}%
The massless fermionic field is conformally invariant in any dimension and
in this case the problem under consideration is conformally related to the
corresponding problem in the Minkowski spacetime with spatial topology $%
\mathrm{R}^{p}\times (\mathrm{S}^{1})^{q}$. Formula (\ref{DelTConf}) is also
obtained from the relation $\Delta _{p+1}\langle T_{i}^{i}\rangle
_{p,q}=a^{-D-1}(\eta )\Delta _{p+1}\langle T_{i}^{i}\rangle _{p,q}^{\mathrm{%
(M)}}$, with $a(\eta )$ being the scale factor. Comparing expression (\ref%
{DelTConf}) with the corresponding formula from \cite{Bell08} for a
conformally coupled massless scalar field, we see that the following
relation takes place: $\Delta _{p+1}\langle T_{k}^{l}\rangle _{p,q}=-N\Delta
_{p+1}\langle T_{k}^{l}\rangle _{p,q}^{\mathrm{(scalar)}}$.

Now we consider the behavior of the vacuum energy-momentum tensor in the
asymptotic regions of the parameters. For small values of the comoving
length $a(\eta )L_{p+1}$ with respect to the dS curvature radius, $a(\eta
)L_{p+1}\ll \alpha $, to the leading order the topological part in the VEV
of the energy-momentum tensor coincides with that for a massless field given
by formula (\ref{DelTConf}). In particular, the topological part of the
vacuum energy density is positive. This limit corresponds to the early
stages of the cosmological evolution, $t\rightarrow -\infty $, and at these
stages the total VEV is dominated by the topological part. In this limit the
back-reaction effects of the topological terms are important and these
effects can change the dynamics essentially (for a discussion of
back-reaction effects from vacuum fluctuations on the dynamics of dS
spacetime see, for instance, \cite{Pere07} and references therein).

For large values of the comoving length, $a(\eta )L_{p+1}\gg \alpha $, the
leading term in the topological part is given by the formula (no summation
over $l$):%
\begin{equation}
\Delta _{p+1}\langle T_{l}^{l}\rangle _{p,q}\approx \frac{NB_{l}\cos
[2mt-2\alpha m\ln (\alpha /L_{p+1})-\phi _{l}]}{2^{p/2}\pi
^{(p+1)/2}V_{q}L_{p+1}^{p+1}\cosh (\alpha m\pi )e^{(D+1)t/\alpha }}.
\label{TllSmall}
\end{equation}%
Here $B_{l}$, $\phi _{l}$ for $l=0,1,\ldots ,p$, are given by relation (\ref%
{B0}), and $B_{l}$, $\phi _{l}$, $l=p+1,\ldots ,D$, are defined by the
formula%
\begin{eqnarray}
B_{l}e^{i\phi _{l}} &=&\frac{-2^{-i\alpha m}}{\Gamma (1/2+i\alpha m)}%
\sum_{n=1}^{\infty }\sum_{n_{p+2}=-\infty }^{+\infty }\cdots
\sum_{n_{D}=-\infty }^{+\infty }\frac{1}{n^{p+2+2i\alpha m}}  \notag \\
&&\times \left[ (p+1+2i\alpha m)f_{p/2+1+i\alpha m}(x)+x^{2}f_{p/2+i\alpha
m}(x)\right] _{x=nL_{p+1}k_{\mathbf{n}_{q-1}}},  \label{Blphipl1}
\end{eqnarray}%
for $l=p+1$, and by the formula
\begin{equation}
B_{l}e^{i\phi _{l}}=\frac{2^{-i\alpha m}L_{p+1}^{2}}{\Gamma (1/2+i\alpha m)}%
\sum_{n=1}^{\infty }\sum_{n_{p+2}=-\infty }^{+\infty }\cdots
\sum_{n_{D}=-\infty }^{+\infty }k_{l}{}^{2}\frac{f_{p/2+i\alpha
m}(nL_{p+1}k_{\mathbf{n}_{q-1}})}{n^{p+2i\alpha m}},  \label{Blphi}
\end{equation}%
for $l=p+2,\ldots ,D$. Formula (\ref{TllSmall}) describes the asymptotic
behavior of the topological part in the late stages of cosmological
evolution corresponding to the limit $t\rightarrow +\infty $. In this limit
the behavior of the topological part for a massive spinor field is damping
oscillatory. As in the case of the fermionic condensate, the damping factor
in the amplitude and the oscillation frequency are the same for all terms in
the sum of Eq. (\ref{TlkdS}) and the total topological term behaves like $%
\langle T_{l}^{k}\rangle _{\mathrm{c}}\propto e^{-(D+1)t/\alpha }\cos \left(
2mt+\phi _{\mathrm{c}}^{\prime }\right) $. As the vacuum energy-momentum
tensor for uncompactified dS spacetime is time-independent, we have similar
damping oscillations in the total energy-momentum tensor $\langle
T_{l}^{k}\rangle _{\mathrm{dS,ren}}+\langle T_{l}^{k}\rangle _{\mathrm{c}}$.
This type of oscillations are absent in the case of a massless field when
the topological parts decay monotonically as $e^{-(D+1)t/\alpha }$. Note
that the topological quantum effects generate an effective potential for the
moduli fields related to the size of extra dimensions. This potential is
proportional to $\Lambda _{\mathrm{eff}}+8\pi G_{D+1}\langle
T_{l}^{k}\rangle _{\mathrm{c}}$. As we have seen, this combination has local
minima in the late stages of cosmological expansion. This provides an
stabilization mechanism in models with a single fermionic field.

The topological parts in the VEV of the energy-momentum tensor for the
fermionic field with antiperiodicity conditions along the compactified
dimensions are investigated in the way similar to that for untwisted field.
The corresponding formulae for the topological parts are obtained from the
expressions for the field with periodicity conditions inserting the factor $%
(-1)^{n}$ in the summation over $n$ and replacing the definition for $k_{%
\mathbf{n}_{q-1}}^{2}$ by (\ref{knq-1Tw}). In particular, for small values
of the comoving length with respect to the dS curvature radius, $a(\eta
)L_{p+1}\ll \alpha $, the topological part of the vacuum energy density is
negative.

\begin{figure}[tbph]
\begin{center}
\begin{tabular}{cc}
\epsfig{figure=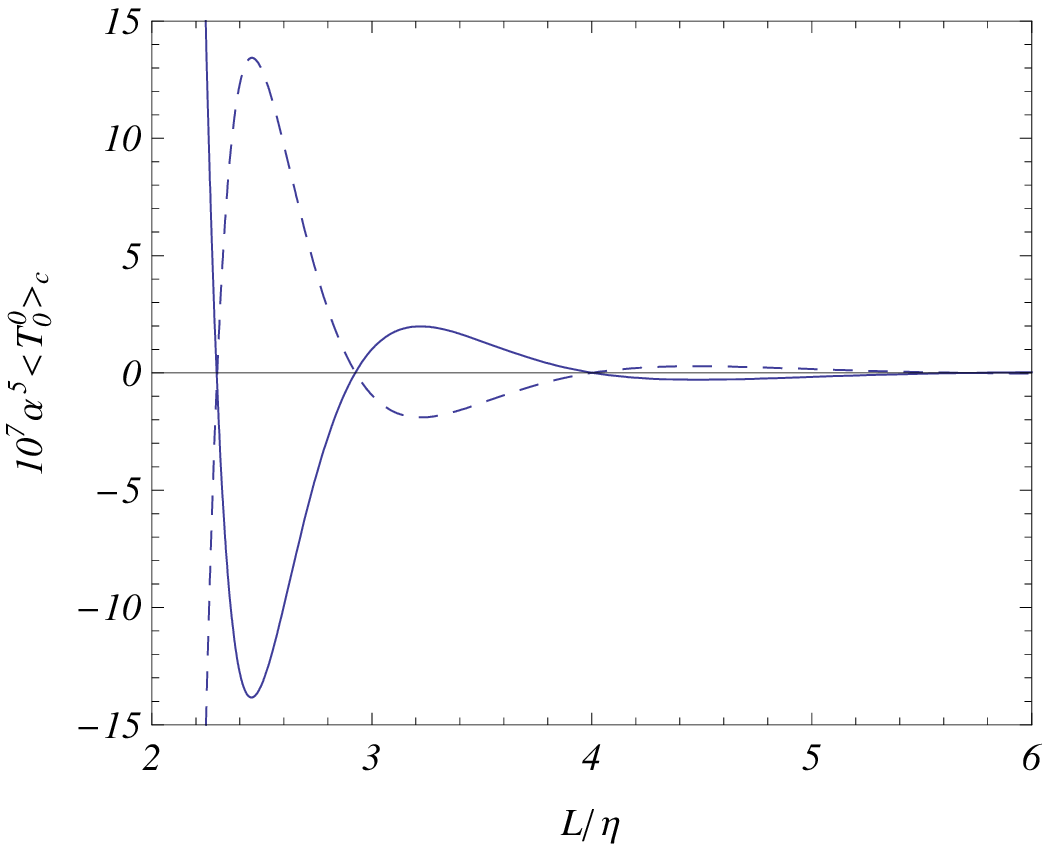,width=7.cm,height=6.cm} & \quad %
\epsfig{figure=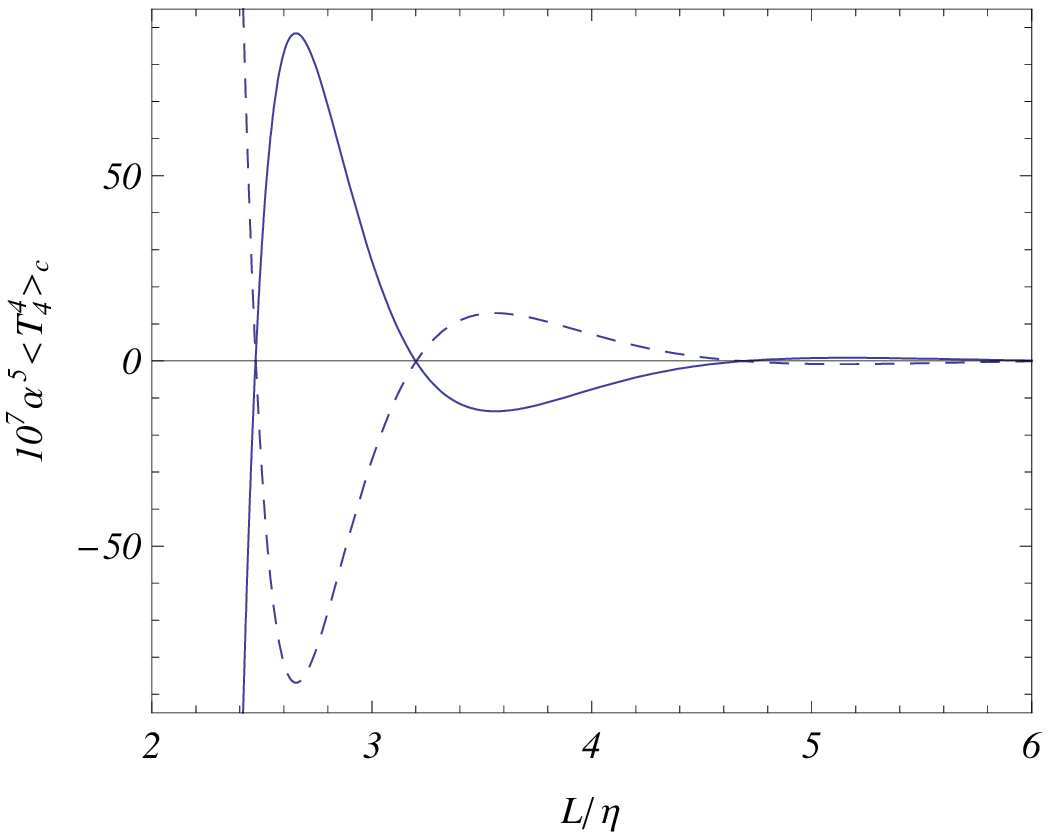,width=7.cm,height=6cm}%
\end{tabular}%
\end{center}
\caption{The topological parts in the VEVs of the energy density (left
panel) and the vacuum stress along the compactified dimension (right panel)
as functions of the ratio $L/\protect\eta $ for the value of the parameter $%
\protect\alpha m=4$ and in the special case of spatial topology $\mathrm{R}%
^{3}\times \mathrm{S}^{1}$. Full/dashed curves correspond to fields with
periodicity/antiperiodicity conditions along the compactified dimension. }
\label{fig2}
\end{figure}

In figure \ref{fig2} we have plotted the topological parts in the VEVs of
the energy density (left panel) and the vacuum stress along the compactified
dimension (right panel) as functions of the ratio $L/\eta $ for the value of
the parameter $\alpha m=4$ and in the special case of $\mathrm{dS}_{5}$
having spatial topology $\mathrm{R}^{3}\times \mathrm{S}^{1}$ with the
length of the compactified dimension $L_{4}=L$. This topology corresponds to
the original Kaluza-Klein model \cite{Kalu21}. Note that the ratio $L/\eta
=a(\eta )L/\alpha $ is the comoving length of the compactified dimension in
units of the dS curvature radius. Full/dashed curves correspond to the
fields with periodicity/antiperiodicity conditions along the compactified
dimension. Note that for the massless fermionic field in dS spacetime with
the spatial topology $\mathrm{R}^{3}\times \mathrm{S}^{1}$ one has%
\begin{equation}
\langle T_{k}^{l}\rangle _{\mathrm{c},m=0}=\frac{b\zeta (5)}{\pi ^{2}}\left(
\frac{\eta }{\alpha L}\right) ^{5}\mathrm{diag}(1,1,1,1,-4),
\label{TlkR2S1m0}
\end{equation}%
where $b=3$ for the field with periodicity condition and $b=-45/16$ for the
field with antiperiodicity condition, $\zeta (x)$ is the Riemann zeta
function. As we have explained before, in the limit $L/\eta \ll 1$, the
tensor (\ref{TlkR2S1m0}) is the leading term in the corresponding asymptotic
expansion for the VEV of the energy-momentum tensor of the massive field.

\begin{figure}[tbph]
\begin{center}
\epsfig{figure=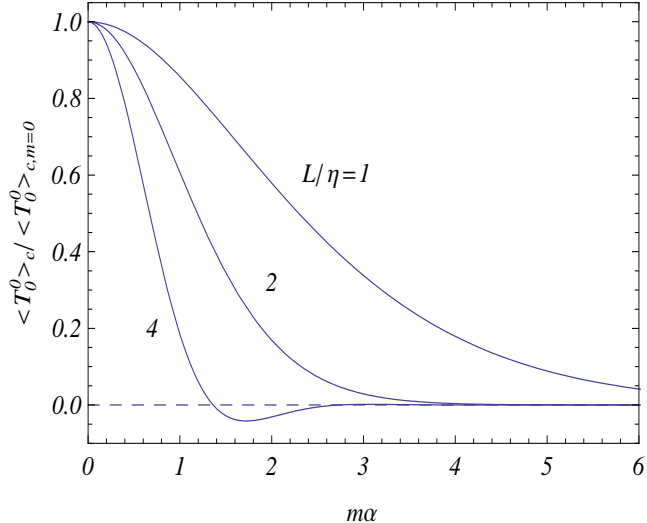,width=8.5cm,height=7cm}
\end{center}
\caption{The ratio $\langle T_{0}^{0}\rangle _{\mathrm{c}}/\langle
T_{0}^{0}\rangle _{\mathrm{c},m=0}$ as a function of the parameter $\protect%
\alpha m$ for a fermionic field with periodicity conditions in $\mathrm{dS}%
_{5}$ with spatial topology $\mathrm{R}^{3}\times \mathrm{S}^{1}$. The
numbers near the curves correspond to the values of the ratio $L/\protect%
\eta $.}
\label{fig3}
\end{figure}

In order to illustrate the dependence of the topological parts on the mass,
in figure \ref{fig3} the ratio $\langle T_{0}^{0}\rangle _{\mathrm{c}%
}/\langle T_{0}^{0}\rangle _{\mathrm{c},m=0}$ is plotted as a function of
the parameter $\alpha m$ for fermionic field with periodicity conditions in $%
\mathrm{dS}_{5}$ with topology $\mathrm{R}^{3}\times \mathrm{S}^{1}$. The
numbers near the curves correspond to the values of the ratio $L/\eta $ with
$L$ being the length of the compactified dimension. According with formula (%
\ref{TllSmall}), for large values of the mass the topological part is an
oscillating function. The amplitudes of these oscillations are exponentially
suppressed and in the given scale they can be seen only for the case $L/\eta
=4$.

The results obtained in the present paper can be used to consider
the role of quantum topological effects in two types of models.
For the first class one has $p=3$, $q\geqslant 1$, and they
correspond to the universe with Kaluza-Klein--type extra
dimensions. For the second type of models $D=3$ and the results
given above describe how the properties of the universe with dS
geometry are changed by one-loop quantum effects induced by the
compactness of spatial dimensions. These effects essentially
depend on the ratio of the dS curvature radius to the Compton
wavelength related to the mass of the field, the parameter $\alpha
m$. In the inflationary period with the characteristic energy $E$,
for this parameter one has $\alpha m\approx 2\times
10^{-15}(m/m_{e})(10^{15}\,\mathrm{GeV}/E)^{2}$, with $m_{e}$
being the mass of the electron. In the standard models of the
inflation with the energy scale of the order
$10^{15}\,\mathrm{GeV}$ the effects due the finite mass are small,
unless the mass is extremely large, and the topological part
in the VEV of the energy-momentum tensor is well-approximated by formula (%
\ref{DelTConf}). For a field with the mass of the order $1\,\mathrm{GeV}$
the effects related to the finite mass are important for the energy scales $%
E\lesssim 10^{9}\,\mathrm{GeV}$. For the recent epoch of the cosmological
expansion with the characteristic energy scale $E\approx 2\times 10^{-3}\,%
\mathrm{eV}$ one has $\alpha m\gg 1$ and for the comoving lengths of the
compactifeied dimensions much less than the horizon scale, $a(t)L_{l}\ll
\alpha $, the effects related to the curvature of the spacetime are small.
In this case formula (\ref{TopTll}) reduces to the corresponding result in
the Minkowski spacetime with topology $\mathrm{R}^{p}\times (\mathrm{S}%
^{1})^{q}$. In the models with $p=3$ and for fields with the mass $%
mL_{l}\lesssim 1$, the topological part of the vacuum energy density is of
the order of the dark energy for the size of the extra dimensions of the
order $10^{-3}\,\mathrm{cm}$ (for a recent discussion see also \cite{Milt03}%
). The models with such large extra dimensions can be realized
within the framework of braneworld scenario. However, it should be
noted that in braneworlds additional contributions to the VEVs are
present due to the imposition of boundary conditions on the branes
(combined effects induced by non-trivial topology and boundary
conditions in braneworld models on AdS bulk are considered in
\cite{Flac03}).

As we have mentioned before at early stages of the cosmological
expansion the separate terms in the topological part for the VEV
of the energy-momentum tensor are approximated by formula
(\ref{DelTConf}). In this
limit the topological part is presented in the form%
\begin{equation}
\langle T_{i}^{k}\rangle _{\mathrm{c}}=a^{-D-1}(t)\langle T_{i}^{k}\rangle _{%
\mathrm{c}}^{\mathrm{(M)}},  \label{Tikcearly}
\end{equation}%
where $\langle T_{i}^{k}\rangle _{\mathrm{c}}^{\mathrm{(M)}}$ is the
corresponding quantity for a conformally coupled massless scalar field in
the Minkowski spacetime with spatial topology $\mathrm{R}^{p}\times (\mathrm{%
S}^{1})^{q}$. Formula (\ref{Tikcearly}) is valid for general case of the
scale factor $a(t)$. This allows us to investigate the back reaction effects
of the topological part on the cosmological evolution in the limit under
consideration. These effects are determined from the $(D+1)$-dimensional
Einstein equations. The corresponding equation for the scale factor has the
form%
\begin{equation}
\frac{1}{a^{2}}\left( \frac{da}{dt}\right) ^{2}=H^{2}+\frac{\beta }{a^{D+1}},
\label{FriedEq}
\end{equation}%
where we have introduced the notations%
\begin{equation}
H^{2}=\frac{2\Lambda _{\mathrm{eff}}}{D(D-1)},\;\beta =\frac{16\pi G_{D+1}}{%
D(D-1)}\langle T_{0}^{0}\rangle _{\mathrm{c}}^{\mathrm{(M)}}.  \label{Hbeta}
\end{equation}%
The solutions of this equation are the functions%
\begin{eqnarray}
a(t) &=&\left[ \sqrt{\beta /H^{2}}\sinh \left( \frac{D+1}{2}Ht\right) \right]
^{2/(D+1)},\;\mathrm{for}\;\beta >0,  \notag \\
a(t) &=&\left[ \sqrt{-\beta /H^{2}}\cosh \left( \frac{D+1}{2}Ht\right) %
\right] ^{2/(D+1)},\;\mathrm{for}\;\beta <0.  \label{a(t)}
\end{eqnarray}%
For $\beta =0$ the solution is given by the line element (\ref{ds2deSit})
with $\alpha =1/H$.

As it is seen from (\ref{a(t)}), in the case $\beta >0$ at $t=0$ one has a
singularity and the size of the universe is zero. Near this singularity the
semiclassical approximation is not applicable. In accordance with (\ref%
{DelTConf}) this case is realized for the fermionic field with periodicity
conditions. For the case $\beta <0$ the scale factor has minimum value $%
a_{\min }=(-\beta /H^{2})^{1/(D+1)}$ and for the topology $(\mathrm{S}%
^{1})^{D}$ we can estimate the probability of creation $P$ of the
corresponding universe in the way similar to that used in
\cite{Zeld84,Gonc85}. Note that this case can be realized by the
fermionic field with antiperiodicity conditions, when the
asymptotic expression for the topological part in the energy
density is given by (\ref{DelTConf}) with an additional factor
$(-1)^{n}$. The
corresponding instanton is obtained from (\ref{a(t)}) by the replacement $%
t\rightarrow -i\tau $:%
\begin{equation}
a_{\mathrm{e}}(\tau )=\left[ \sqrt{-\beta /H^{2}}\cos \left( \frac{D+1}{2}%
H\tau \right) \right] ^{2/(D+1)},  \label{aetau}
\end{equation}%
with $\tau $ being the euclidean time and $H|\tau |\leqslant \pi /(D+1)$.
The euclidean action for the instanton is given by the expression%
\begin{equation}
S_{\mathrm{e}}=iA_{D}V_{D}|\langle T_{0}^{0}\rangle _{\mathrm{c}}^{\mathrm{%
(M)}}|^{D/(D+1)},  \label{Se}
\end{equation}%
with the notation%
\begin{equation}
A_{D}=\frac{4\sqrt{\pi }}{D+1}\frac{\Gamma \left( \frac{D-1}{2(D+1)}\right)
[D(D-1)]^{-D/(D+1)}}{\Gamma \left( \frac{D}{D+1}\right) (16\pi
G_{D+1}H^{D-1})^{1/(D+1)}}.  \label{AD}
\end{equation}
For the creation probability one has $P\sim \exp (-2|S_{\mathrm{e}}|)$. In
the case $D=3$ this formula for the creation probability reduces to the one
discussed in \cite{Zeld84,Gonc85}.

For the further discussion we note that when the sizes of the
compactified dimensions are the same, $L_{p+1}=\cdots =L_{D}=L$,
the topological part in
the energy density of the twisted fermionic field for the topology $\mathrm{R%
}^{p}\times (\mathrm{S}^{1})^{D-p}$ is presented in the form%
\begin{equation}
\langle T_{0}^{0}\rangle _{\mathrm{c},\mathrm{R}^{p}\times (\mathrm{S}%
^{1})^{D-p}}^{\mathrm{(M)}}=\frac{a_{D,p}}{L^{D+1}},  \label{T00L}
\end{equation}%
where, in accordance with (\ref{DelTConf}), the coefficient is given by the
expression%
\begin{equation}
a_{D,p}=2N\sum_{l=p}^{D-1}\sum_{n=1}^{\infty }\sum_{n_{l+2}=-\infty
}^{+\infty }\cdots \sum_{n_{D}=-\infty }^{+\infty }\frac{f_{l/2+1}(\pi n%
\sqrt{(2n_{l+2}+1)^{2}+\cdots +(2n_{D}+1)^{2}})}{(-1)^{n}(2\pi n^{2})^{l/2+1}%
}.  \label{aDp}
\end{equation}%
Now let us consider the creation probability in two special cases. For the
first one we assume the topology $(\mathrm{S}^{1})^{D}$ with $L_{1}=\cdots
=L_{D}=L$. In this case for the creation probability one has%
\begin{equation}
P\sim \exp [-2A_{D}|a_{D,0}|^{D/(D+1)}].  \label{P1}
\end{equation}%
In the second case we assume the topology $(\mathrm{S}^{1})^{D}$ with $%
L_{1}=\cdots =L_{p}=L_{\mathrm{l}}$, $L_{p+1}=\cdots
=L_{D}=L_{\mathrm{s}}$ and $L_{\mathrm{l}}\gg L_{\mathrm{s}}$. For
this case, to the leading order one has $\langle T_{0}^{0}\rangle
_{\mathrm{c}}^{\mathrm{(M)}}\approx
a_{D,p}/L_{\mathrm{s}}^{D+1}$ and the creation probability takes the form%
\begin{equation}
P\sim \exp [-2A_{D}|a_{D,p}|^{D/(D+1)}(L_{\mathrm{l}}/L_{\mathrm{s}})^{p}].
\label{P2}
\end{equation}

\section{Conclusion}

\label{sec:Conc}

In the present paper we have investigated the fermionic condensate and the
VEV of the energy-momentum tensor for a massive fermionic field in
higher-dimensional dS spacetime with toroidally compactified spatial
dimensions. In Section \ref{sec:UncompdS} we have considered the
corresponding quantities in uncompactified odd-dimensional dS spacetime
assuming that the field is prepared in the Bunch-Davies vacuum state. The
renormalization is done by using the dimensional regularization procedure.
Closed expressions, formulae (\ref{FermConddSRen}) and (\ref{TkldSren}), are
derived for the renormalized fermionic condensate and the VEV of the
energy-momentum tensor respectively. For large values of the mass these
quantities are exponentially suppressed. Note that in even-dimensional dS
spacetime for large mass the suppression is power-law.

Further, we have investigated one-loop quantum effects on the fermionic
vacuum induced by the non-trivial topology of spatial dimensions.
Specifically, we have considered the dS spacetime with toroidally
compactified dimensions having the spatial topology $\mathrm{R}^{p}\times (%
\mathrm{S}^{1})^{q}$. For the evaluation of the vacuum densities, the
mode-summation procedure is employed. In this procedure we need to know the
corresponding eigenspinors satisfying appropriate boundary conditions along
the compactified dimensions. These eigenspinors are constructed in section %
\ref{sec:EigFunc} for both fields obeying periodicity and antiperiodicity
boundary conditions. By using these eigenfunctions and applying to the
mode-sums the Abel-Plana formula, the VEVs for the spatial topology $\mathrm{%
R}^{p}\times (\mathrm{S}^{1})^{q}$ are presented in the form of the sum of
the corresponding quantity in the topology $\mathrm{R}^{p+1}\times (\mathrm{S%
}^{1})^{q-1}$ and of the part which is induced by the compactness of $(p+1)$%
th dimension. For fields obeying periodicity conditions, the topological
parts are given by formulae (\ref{DeltCond2}) and (\ref{TopTll}) for
fermionic condensate and energy-momentum tensor, respectively. The
corresponding formulae for the field with antiperiodicity conditions are
obtained from those for the field obeying periodicity conditions inserting
the factor $(-1)^{n}$ in the summation over $n$ and replacing the definition
for $k_{\mathbf{n}_{q-1}}^{2}$ by (\ref{knq-1Tw}).

The topological parts are finite and the renormalization procedure is needed
only for the uncompactified dS spacetime. These parts are time-dependent and
break the dS symmetry. The corresponding vacuum stresses along the
uncompactified dimensions coincide with the energy density and, hence, in
the uncompactified subspace the equation of state for the topological part
of the energy-momentum tensor is of the cosmological constant type. For a
massless fermionic field the problem under consideration is conformally
related to the corresponding problem in the Minkowski spacetime with spatial
topology $\mathrm{R}^{p}\times (\mathrm{S}^{1})^{q}$ and the topological
part of the fermionic condensate vanishes. For the VEV of the
energy-momentum tensor we have the standard relation $\langle
T_{k}^{l}\rangle _{c}=a^{-(D+1)}(\eta )\langle T_{k}^{l}\rangle _{c}^{%
\mathrm{(M)}}$ between the topological contributions.

For a massive fermionic field, in the limit when the comoving length of a
compactified dimension is much smaller than the dS curvature radius, the
topological part in the VEV of the energy-momentum tensor coincides with the
corresponding quantity for a massless field and is conformally related to
the VEV in toroidally compactified Minkowski spacetime. In particular, the
topological part in the vacuum energy density is positive for an untwisted
fermionic field. This limit corresponds to the early stages of the
cosmological evolution and the topological parts dominate over the
uncompactified dS parts. At these stages the back-reaction effects of the
topological terms are important and these effects can essentially change the
dynamics of the model. In the opposite limit, when the comoving lengths of
the compactified dimensions are large with respect to the dS curvature
radius, in the case of a massive field the asymptotic behavior of the
topological parts are oscillatory damping for both fermionic condensate and
the energy-momentum tensor and their respectively leading term are given by
formulae (\ref{CondLate}) and (\ref{TllSmall}). These formulae describe the
behavior of the topological parts in the late stages of the cosmological
expansion. As the corresponding uncompactified dS parts are
time-independent, we have similar oscillations in the total VEVs as well.
Note that this type of oscillatory behavior is absent for a massless
fermionic field.

\section*{Acknowledgments}

E.R.B.M. thanks Conselho Nacional de Desenvolvimento Cient\'{\i}fico e Tecnol%
\'{o}gico (CNPq) for partial financial support, FAPESQ-PB/CNPq (PRONEX) and
FAPES-ES/CNPq (PRONEX). A.A.S. was supported by the Armenian Ministry of
Education and Science Grant No. 119 and by Conselho Nacional de
Desenvolvimento Cient\'{\i}fico e Tecnol\'{o}gico (CNPq).

\end{document}